\documentclass[aps,amsmath,amssymb,floats]{revtex4}

\usepackage{graphicx}  

\begin{document}

\title{Normal Emission Study on Body-Centered-Cubic Ni\footnote[2]{As a supplemental material
for C. S. Tian, D. Qian, D. Wu, R. H. He, Y. Z. Wu {\sl et al.},
Phys. Rev. Lett. {\bf 94}, 137210 (2005).}}

\author{R. H. He\footnote[1]{Electronic mail: ruihuahe@stanford.edu; Present address:
Department of Applied Physics, Stanford University, Stanford,
California 94305}, C. S. Tian, D. Qian, D. Wu, Y. Z. Wu, W. X.
Tang, L. F. Yin, Y. S. Shi, G. S. Dong, and X. F. Jin}

\address{Physics Department and Applied Surface Physics
State Key Laboratory, Fudan University, Shanghai 200433, China}

\begin{abstract}
The first normal emission experiment result on bcc Ni ultrathin
film is presented in comparison with the one on fcc Ni. Its
agreement with band structure calculation, by supplementing our
former results on the magnetic properties of bcc Ni, verifies from
the electronic viewpoint the successful epitaxy of this metastable
phase of Ni which doesn't exist in nature.
\end{abstract}

\pacs{PACS numbers: 75.30.¨Cm, 79.60.-i, 81.20.¨Cn}

\maketitle

\maketitle

The different crystalline structure of bcc Ni from that of its fcc
counterpart implies a distinct electronic structure should be
readily observable. In order to verify this, the angle-resolved
photoemission study with variable photon energies (h$\nu$'s) was
performed on 1.6-nm thick bcc Ni and fcc Ni films, respectively,
along the $<001>$ direction (normal emission), with fcc Ni film
grown on Cu(001) substrate for comparison. The two films at the
chosen thickness are both expected to have well-defined
crystalline structures with pretty good quality thus making
possible the probe of their intrinsic line shapes basically free
of the contribution from the substrates at the chosen film
thickness. Data were obtained at ~170K that is below the $T_c$'s
of the Ni films with an incident angle of radiation $55^\circ$.

\begin{figure}[b]
\includegraphics[width=3.5in]{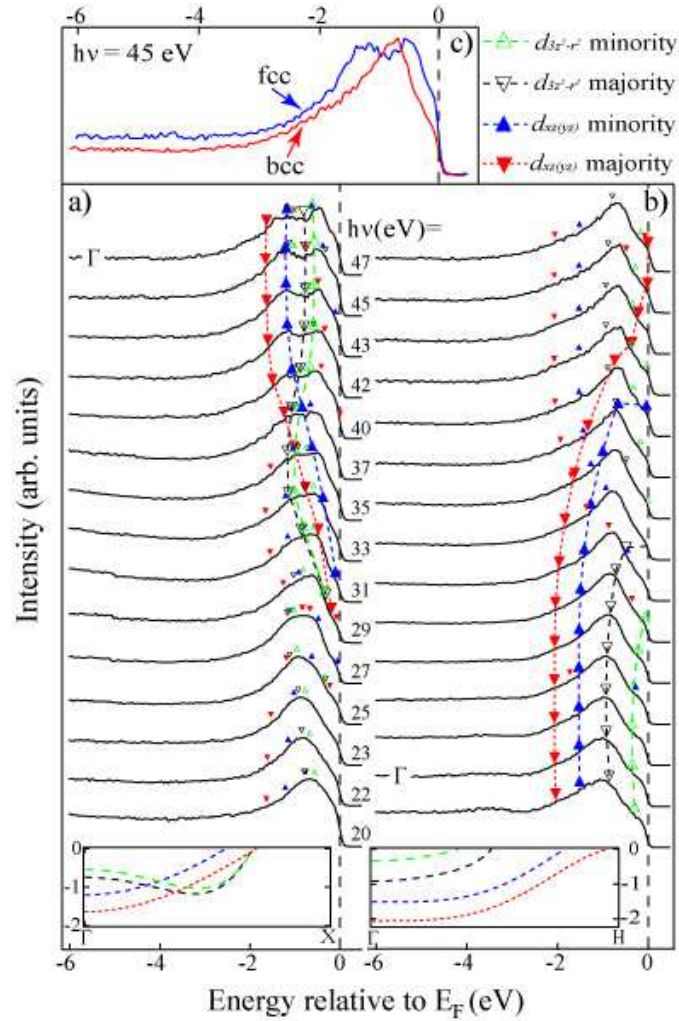}\vspace{0cm}
\caption{(color) Normal emission spectra on (a) fcc Ni/Cu(001),
(b) bcc Ni/GaAs(001) are shown. Insets: the theoretical
dispersions of the spin-resolved initial states involved in the
direct transitions are shown. The theory predicted band maximums
are appended on the EDC overlay. Features resulted from the
transition in the first in-plane Brillouin zone (BZ) are specified
using larger symbols along with the dash guidelines to the eye.
(c) a selected contrast of the normalized spectra taken with
h$\nu$=45 eV.}\label{NiNormalEmission}
\end{figure}

As shown in Fig. \ref{NiNormalEmission}, two main features can be
identified by the eye showing a great difference between the
valence band structures of the two crystalline phases. First, a
contrast in line shape is found for each pair of energy
distribution curves (EDC's) obtained at the same $h\nu$ on the two
crystals, i.e., a multi-component nature is more readily observed
in the fcc Ni spectra while a seemingly single broad peak persists
in the whole $h\nu$ range in the bcc case. Second, in contrast to
the apparent band dispersion in the fcc case, the valence bands of
bcc Ni conjure to exhibit a virtually non-dispersive character
other than a discernible Fermi crossing of one or two of the
low-lying states.

In order to obtain further insights into the dispersion along
$<001>$, identifications of the experimental peak features based
on the first principle calculation results are made. Instead of
following a widely-used scheme where points with each representing
an experimental peak feature are sprinkled over the predicted
dispersion, we prefer a straightforward expression by
superimposing the theoretical guidelines to the experimental EDC's
overlay. This bypasses the difficulty in achieving an objective
determination on the positions of the overlapping peaks from a
single broad feature in the experimental curves and thus increases
the robustness of comparison.

In Fig. \ref{NiNormalEmission}, under the normal emission
selection rule, the theoretical band maximums given by a routine
conversion\cite{ConversionNote} of the G-DFT\cite{GDFT} as well as
the LAPW\cite{LAPW} dispersion results are indicated respectively
for the fcc and bcc case. Note that the mini symbols on the EDC's
represent features by transitions beyond the first in-plane
BZ\cite{LuThesis} and account for the features in the EDC's
($h\nu<$27 eV) where no direct transition in the first in-plane BZ
within the interested energy window exists. A nice agreement
between the theory and the experiment is obtained on the main EDC
features and their dispersions as well as the persistent Fermi
cutoffs observed due to the sequential Fermi crossings of several
bands obscured by the finite momentum resolution\cite{ZXTiTe2}.

In the fcc Ni spectra, two main peaks are observed around the BZ
center ($\Gamma$ point), corresponding to the EDC taken with
h$\nu$=45 eV\cite{fccNi_uv, fccNisurfacestate}. In contrast, for
the bcc case, a persisting broad line shape composed of flat bands
in parallel dispersions as well as the gradual dying out of the
$d_{3z^2-r^2}$ minority band starting from 0.3 eV toward $E_F$
suggest that $h\nu$=22 eV probes the transitions around $\Gamma$
point. The quite opposite correspondence between the $h\nu$'s and
the momentum fraction along $<001>$ is caused by the difference in
lattice constant between bcc and fcc Ni.

\section{Appendix}

\begin{figure}[t]
\includegraphics[width=3in]{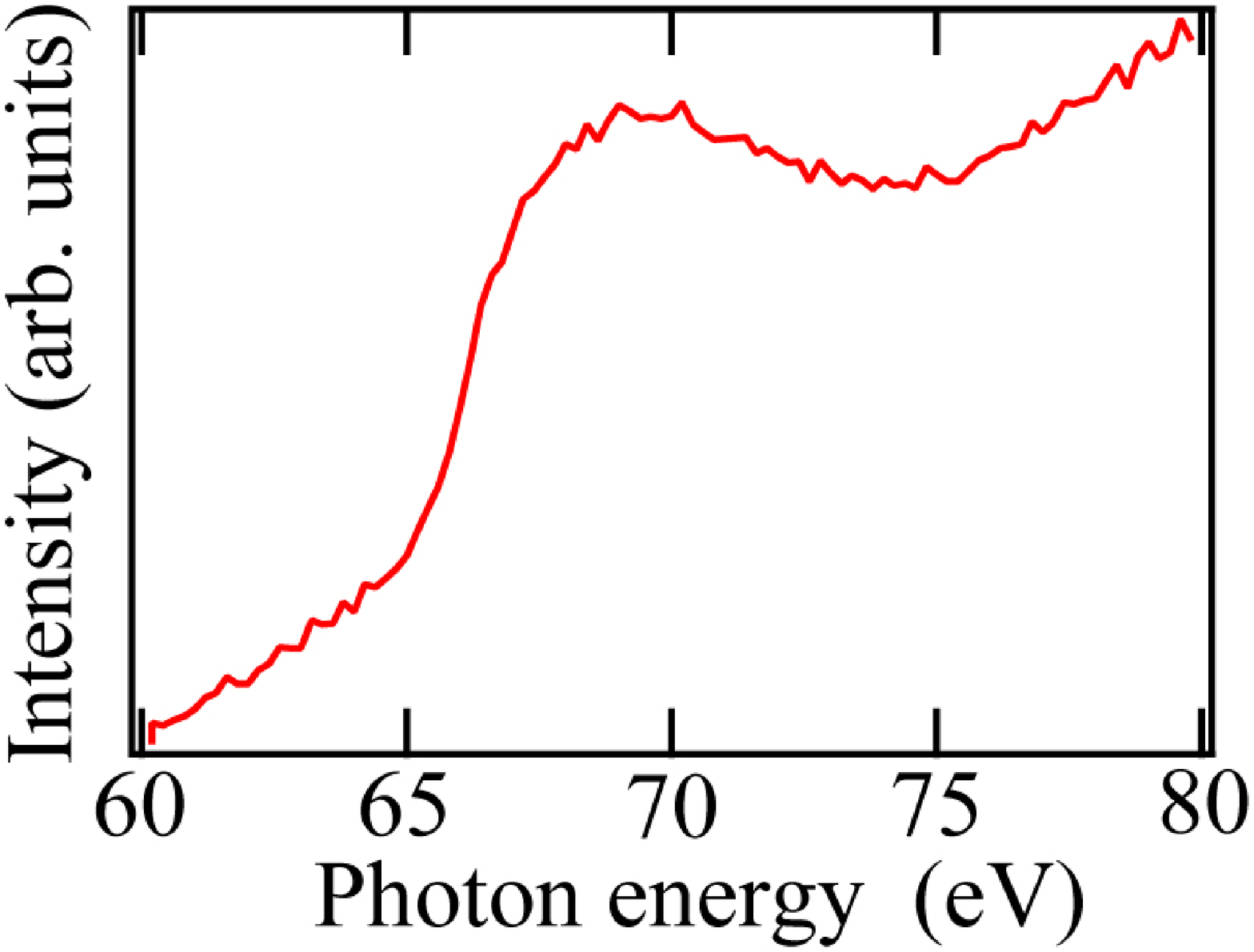}\vspace{0cm}
\caption{(color) The Constant Initial State measurement on the
6-eV resonance feature.}\label{6eVCIS}
\end{figure}

We also find the resonance of the 6-eV satellite peak with the
68-eV photon in bcc Ni/GaAs(001) (Fig. \ref{6eVCIS}) by the
Constant Initial State measurement on the 6-eV feature). This
observation points to its structural independence confirming that
this many-body phenomenon originates from the atomic nature of Ni
(consistent with Ref. \cite{6eVresonance}). The interest of this
conclusion is related to the correlation effect issue in bcc Ni.
But since it occurs both in fcc and bcc Ni, our manuscript which
focuses on the fcc-bcc spectral difference doesn't include a
detailed discussion of this result.


\begin{thebibliography}{99}
\bibitem{ConversionNote} A free-electron-like (FEL) final state is
assumed. The inner potential chosen for fcc final state is -5 eV,
which is consistent with Ref. \cite{{fccNi_uv}}. The 3 eV for bcc
later is determined by achieving a qualitatively best fit to the
main EDC features and consistent with the result of a direct
parabolic fit on the $\Delta_1$ conduction band.
\end{thebibliography}
\end{document}